\documentclass[aps,prb,twocolumn,groupedaddress]{revtex4}

\usepackage{graphicx,amsmath}

\begin{document}

\title{Tracking the effects of interactions on spinons in gapless Heisenberg chains}
\author{Jean-S{\'e}bastien Caux$^1$, Hitoshi Konno$^2$, Mark Sorrell$^3$ and Robert Weston$^4$}

\affiliation{$^1$Institute for Theoretical Physics, Universiteit van Amsterdam,
Science Park 904, Amsterdam, The Netherlands \\
$^2$Department of Mathematics, Hiroshima University, Higashi-Hiroshima 739-8521, Japan \\
$^3$Department of Mathematics and Statistics, The University of Melbourne, Parkville VIC 3010, Australia \\
$^4$Department of Mathematics, Heriot-Watt University, Edinburgh EH14 4AS, UK.}

\begin{abstract}
We consider the effects of interactions on spinon excitations in Heisenberg spin-$1/2$ chains.
We compute the exact two-spinon part of the longitudinal structure factor of the 
infinite chain in zero field for all values of anisotropy in the gapless 
antiferromagnetic regime, via an exact algebraic approach.
Our results allow us to quantitatively describe the behaviour of these fundamental excitations
throughout the observable continuum,  
for cases ranging from free to fully coupled chains, thereby explicitly mapping the effects 
of `turning on the interactions' in a strongly-correlated system.
\end{abstract}

\maketitle

Interactions in one-dimensional (1d) systems are known to overwhelm constituent
particles, leading to a collective quantum liquid state with low-energy 
excitations 
described by the theory of Tomonaga-Luttinger liquids
\cite{1981_Haldane_JPC_14}.
While the `universal' physics of 1d systems is phenomenologically well understood \cite{GiamarchiBOOK}, 
it almost always remains impossible to precisely track the effects
of `turning on the interactions' on the constituent particles, as one does for
Fermi liquids \cite{1957_Landau_JETP_3} (where
bare fermions are adiabatically connected to Landau quasiparticles).
In this respect, our general understanding of 1d systems can benefit from
nonperturbative solutions of microscopic models, a fundamental example being the 
Heisenberg spin-$1/2$ anisotropic chain \cite{1928_Heisenberg_ZP_49,1958_Orbach_PR_112},
whose Hamiltonian is
\begin{equation}
H = J \sum_{j=1}^N \left(S^x_j S^x_{j+1} + S^y_j S^y_{j+1} + \Delta S^z_j S^z_{j+1}\right).
\label{eq:HXXZ}
\end{equation}
This system is a Tomogana-Luttinger liquid for anisotropy values 
$\Delta$ in the range $-1 < \Delta \leq 1$ (in zero field, with $J > 0$).
Its fundamental excitations
are spinons \cite{1981_Faddeev_PLA_85}: spin-$1/2$ fractionalized objects which can be viewed as
domain walls dressed by quantum fluctuations.  

A way to probe the nature of excitations is to determine how they carry observable correlations, 
an interesting example here being the longitudinal structure factor
\begin{equation}
S^{zz}(k,\omega) \!=\! \frac{1}{N} \!\sum_{j, j'} e^{-ik (j - j')} \!\!\int_{-\infty}^{\infty} \!dt e^{i\omega t} 
\langle S^z_j (t) S^z_{j'} (0) \rangle.
\label{eq:SZZ}
\end{equation}
At $\Delta = 0$, this can be written as a density correlator of Jordan-Wigner fermions.
Only single particle-hole excitations contribute, the exact structure factor being proportional to
their density of states.  For $\Delta > 0$, this picture breaks down \cite{1981_Mueller_PRB_24,1981_Mueller_JPC_14} due to 
nonperturbative effects.

It is the purpose of this paper to track in detail the effects of `turning on' interactions
on the spinon quasiparticles and their ability to carry correlations, 
throughout the gapless antiferromagnetic regime $0 \leq \Delta \leq 1$.
Systems in this regime can be realized and studied experimentally
(for fixed anisotropy) in spin ladder compounds \cite{1998_Totsuka_PRB_57,2001_Watson_PRL_86,2009_Thielemann_PRL_102}
or (in principle for generic anisotropy) using optical lattices \cite{2003_Duan_PRL_91}.
Focusing on zero temperature, we will compute the exact two-spinon contribution to (\ref{eq:SZZ})
directly in the thermodynamic limit $N \rightarrow \infty$,
using an adaptation of the `vertex operator approach' \cite{JimboBOOK}. 
Our results provide a strict lower bound and (for practical purposes) an
extremely accurate representation for the complete correlator 
of the infinite system (more that 99\% for anisotropies below $0.5$) throughout the observable excitation continuum.
They provide a robust benchmark for assessing the lineshapes
obtained for finite systems directly from integrability \cite{2005_Caux_PRL_95,2005_Caux_JSTAT_P09003}
or using variants of the density matrix renormalization group (DMRG) \cite{2004_White_PRL_93,2006_Sirker_PRB_73} 
or quantum Monte Carlo (QMC) \cite{2008_Syljuasen_PRB_78},
and confirming the threshold behaviour predicted using 
field theory \cite{2006_Pustilnik_PRL_96,2006_Pereira_PRL_96,2008_Pereira_PRL_100,2008_Cheianov_PRL_100,2009_Pereira_PRB_79},
complementing it with exact prefactors.

The vertex operator approach was originally developed for $\Delta\geq 1$ where
the Hamiltonian commutes with the action of the quantum group $U_q(\widehat{sl}_2)$.
The representation theory of this quantum group
leads to explicit expressions for states, physical operators and their matrix elements \cite{JimboBOOK}, 
providing building blocks for 
correlations in terms of contributions from intermediate states made of increasing numbers of 
pairs of spinons,
$S^{zz} (k, \omega) = \sum_{m=1}^\infty S^{zz}_{(2m)} (k, \omega)$.  
The calculation of (\ref{eq:SZZ}) 
was treated using the vertex operator approach 
at $\Delta = 1$ for two \cite{1996_Bougourzi_PRB_54,1997_Karbach_PRB_55}
and four spinons \cite{1997_Abada_NPB_497,2006_Caux_JSTAT_P12013},
the combination 
being shown to yield about 99\% overall accuracy. 
The $\Delta > 1$ regime 
was also considered \cite{1998_Bougourzi_PRB_57,2008_Caux_JSTAT_P08006}. 
The physically more interesting quantum critical gapless regime ($0\leq \Delta\leq 1$) remains however 
largely unexplored by these exact thermodynamic methods. Our paper aims to fill this gap.

{\it Spinon excitations --}
The ground state of the gapless $XXZ$ antiferromagnet supports 
spinon excitations \cite{1981_Faddeev_PLA_85} with exact zero-field dispersion relation
$e(p) = v_F | \sin p |$, $p \in [-\pi, 0]$,
where the Fermi velocity is 
$v_F (\Delta) = \frac{\pi J}{2} \frac{\sqrt{1 - \Delta^2}}{\mbox{acos} \Delta}.$
Spinons always appear in pairs, so the simplest 
states which contribute to the structure factor are made of 2 spinons.  
Parametrizing their momentum by $p_1$ and $p_2$, momentum and energy conservation impose
$k = -p_1 - p_2,$
$\omega = e(p_1) + e(p_2).$
The two-spinon states thus form a continuum in $k$-$\omega$ defined by lower and upper boundaries
\begin{equation}
\omega_{2,l} (k) = v_F |\sin k|, \hspace{1cm} \omega_{2,u} (k) = 2 v_F \sin k/2.
\label{eq:boundaries}
\end{equation}

{\it Matrix elements via vertex operator approach --}
The vertex operator approach is also applicable, albeit indirectly, to the gapless region $0 \leq \Delta\leq 1$. 
The strategy \cite{1996_Jimbo_JPA_29,1997_Jimbo_Proceedings_Ascona} is to first generalize the problem to the completely anisotropic Heisenberg model 
$\sum\limits_{j} (J_x S_j^x S_{j+1}^x + J_y S_j^y S_{j+1}^y +J_z S_j^z S_{j+1}^z)$ in the so called principal 
regime $|J_y|\leq J_x\leq J_z$ \cite{BaxterBOOK} for which matrix elements of local operators between the 
vacuum and excited states can be computed exactly using a variant of the vertex operator approach 
\cite{1998_Lashkevich_NPB_516,1998_Lashkevich_JETPL_68,2002_Lashkevich_NPB_621,2005_Kojima_NPB_720}. 
These results can then be mapped to the disordered regime $|J_z|\leq J_y \leq J_x$ \cite{1997_Jimbo_Proceedings_Ascona,2003_Lukyanov_NPB_654} 
before taking the $J_x \rightarrow J_y$ limit to reconstruct the matrix elements for the gapless Hamiltonian (1) with $0\leq \Delta\leq 1$.
In this way we find\cite{CauxKonnoSorrellWeston_TBP} the following exact expression for the two-spinon contribution to $S^{zz}(k,w)$:
\begin{eqnarray}
S^{zz}_2 (k, \omega) = 
\frac{\Theta (\omega_{2,u} (k) - \omega) \Theta (\omega - \omega_{2,l}(k))}{\sqrt{\omega_{2,u}^2(k) - \omega^2}}
\nonumber \\
\times (1 + 1/\xi)^2 \frac{e^{-I_\xi (\rho(k,\omega))}}{\cosh \frac{2\pi \rho(k,\omega)}{\xi} + \cos \frac{\pi}{\xi}}
\label{S_2}
\end{eqnarray}
in which $\xi = \frac{\pi}{\mbox{acos} \Delta} - 1$, 
$\Theta$ is the Heaviside function, and
\begin{equation}
I_{\xi} (\rho) \equiv \int_0^{\infty} \frac{dt}{t} \frac{\sinh (\xi + 1)t}{\sinh \xi t} \frac{\cosh (2t) \cos (4\rho t) - 1}{\cosh t \sinh (2t)}
\label{I_rho}
\end{equation}
in which the parameter $\rho$ is defined as
\begin{equation}
\cosh (\pi \rho(k, \omega)) = \sqrt{\frac{\omega_{2,u}^2(k) - \omega_{2,l}^2(k)}{\omega^2 - \omega_{2,l}^2(k)}}.
\label{rhodef}
\end{equation}

\begin{figure*}
\begin{tabular}{ccc}
\includegraphics[width=75mm]{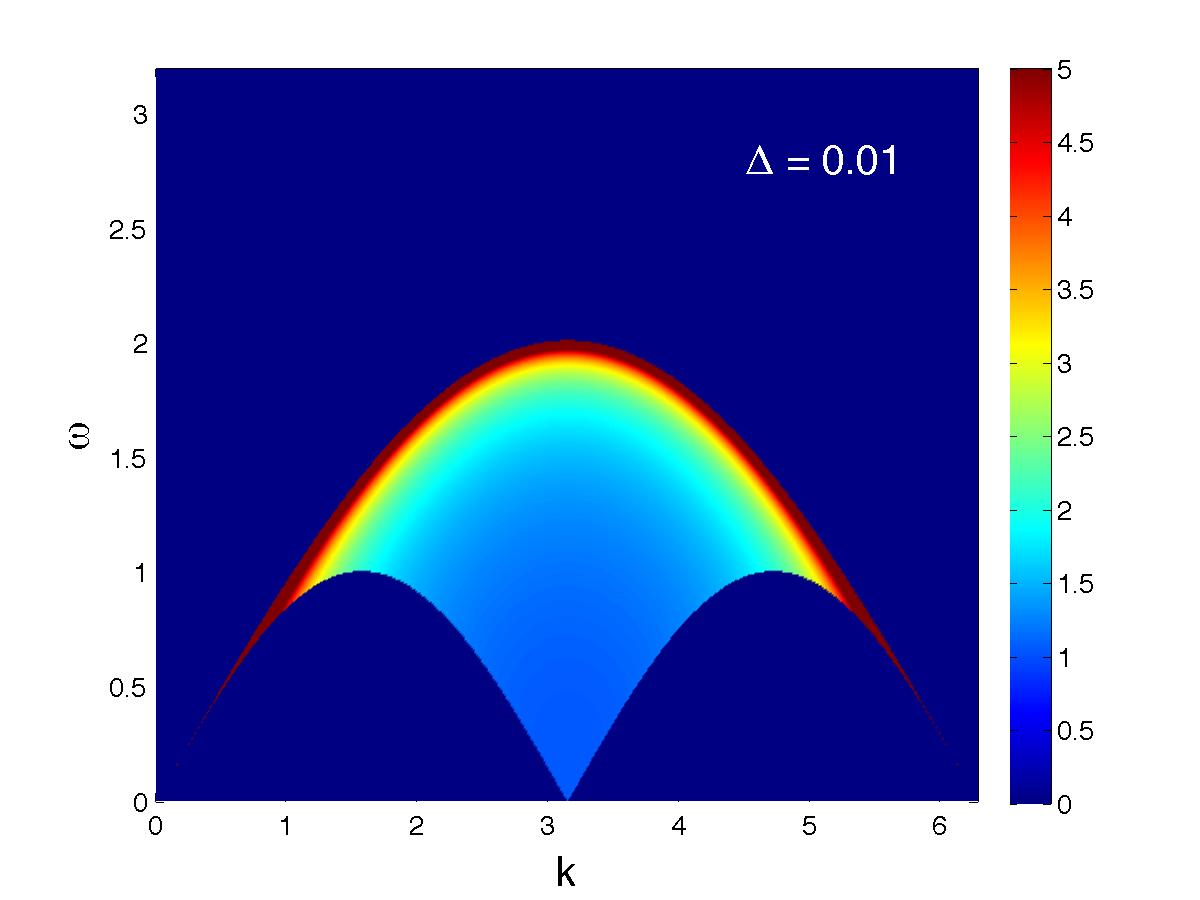}
&
\includegraphics[width=75mm]{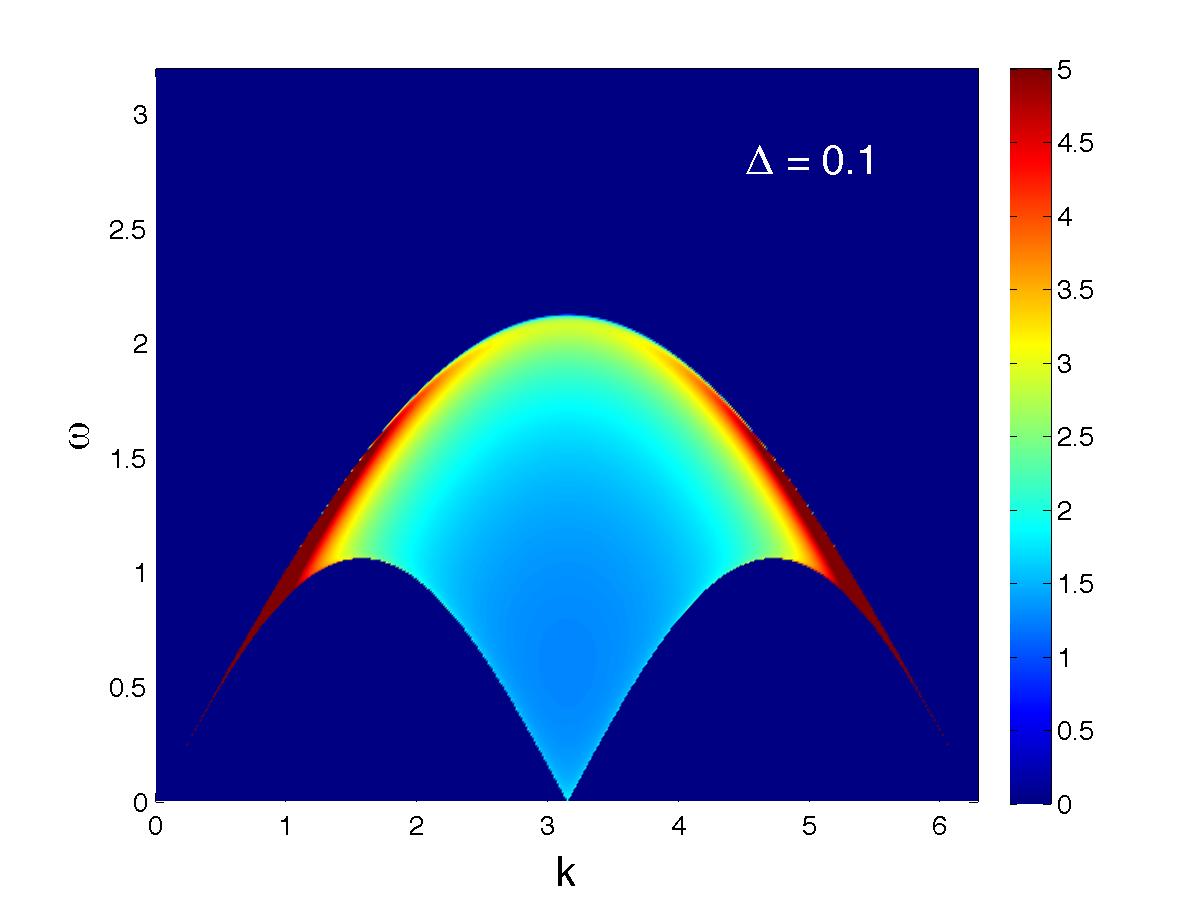}
\\
\includegraphics[width=75mm]{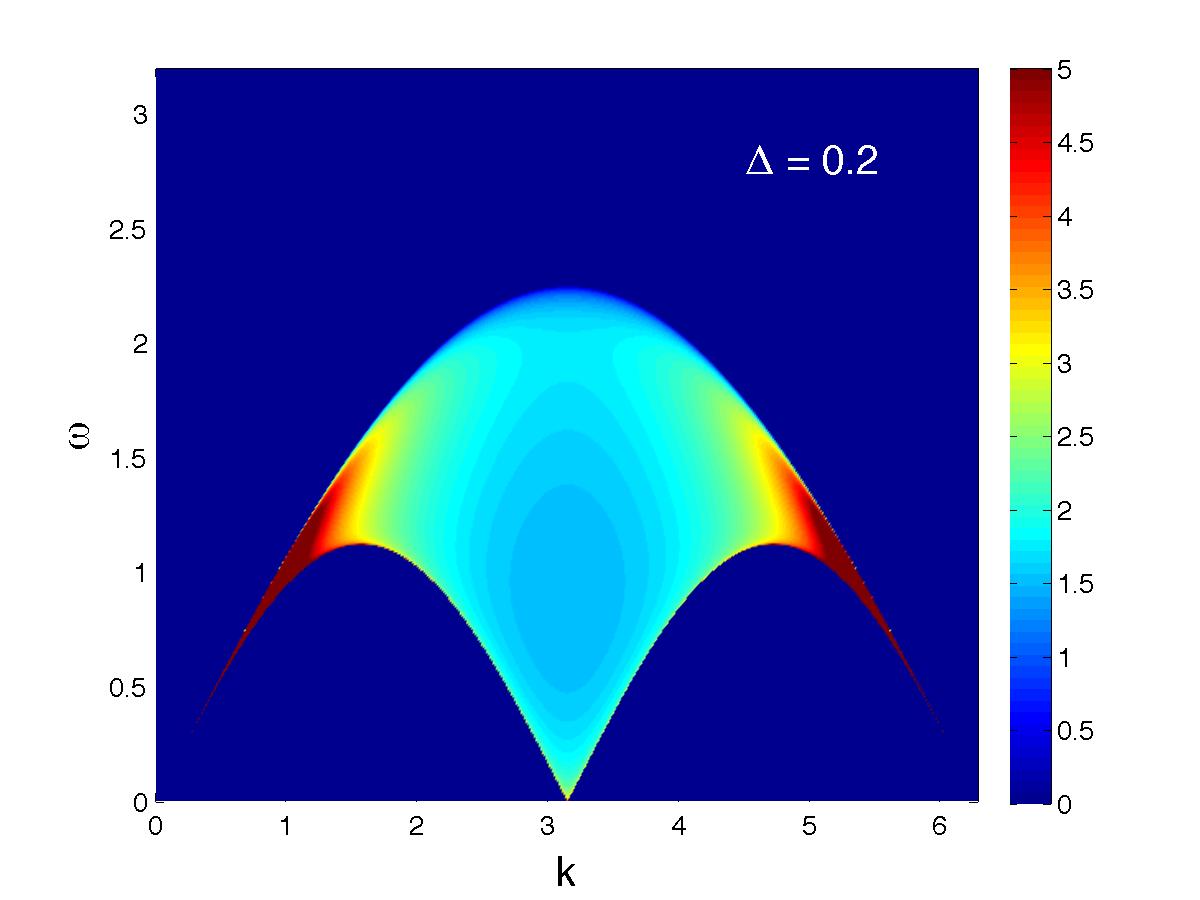}
&
\includegraphics[width=75mm]{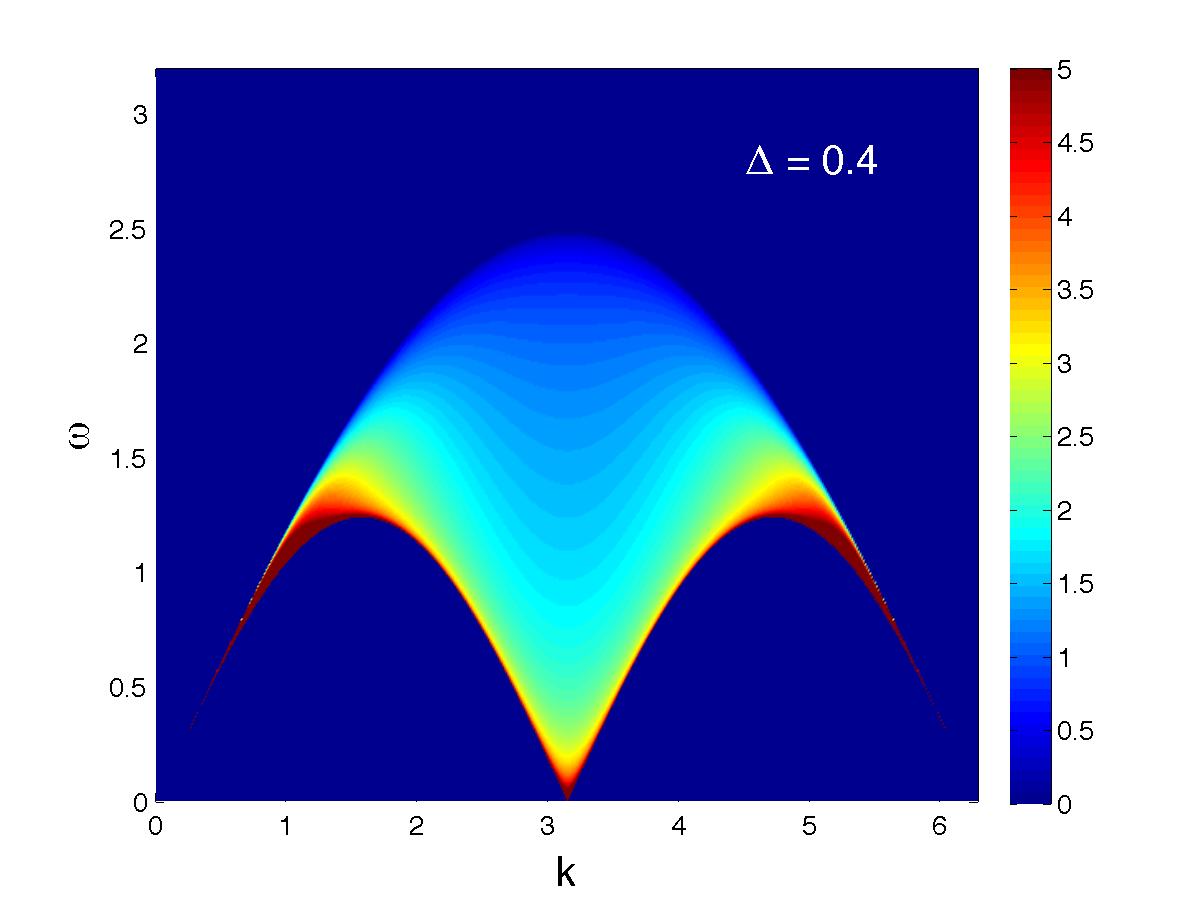}
\\
\includegraphics[width=75mm]{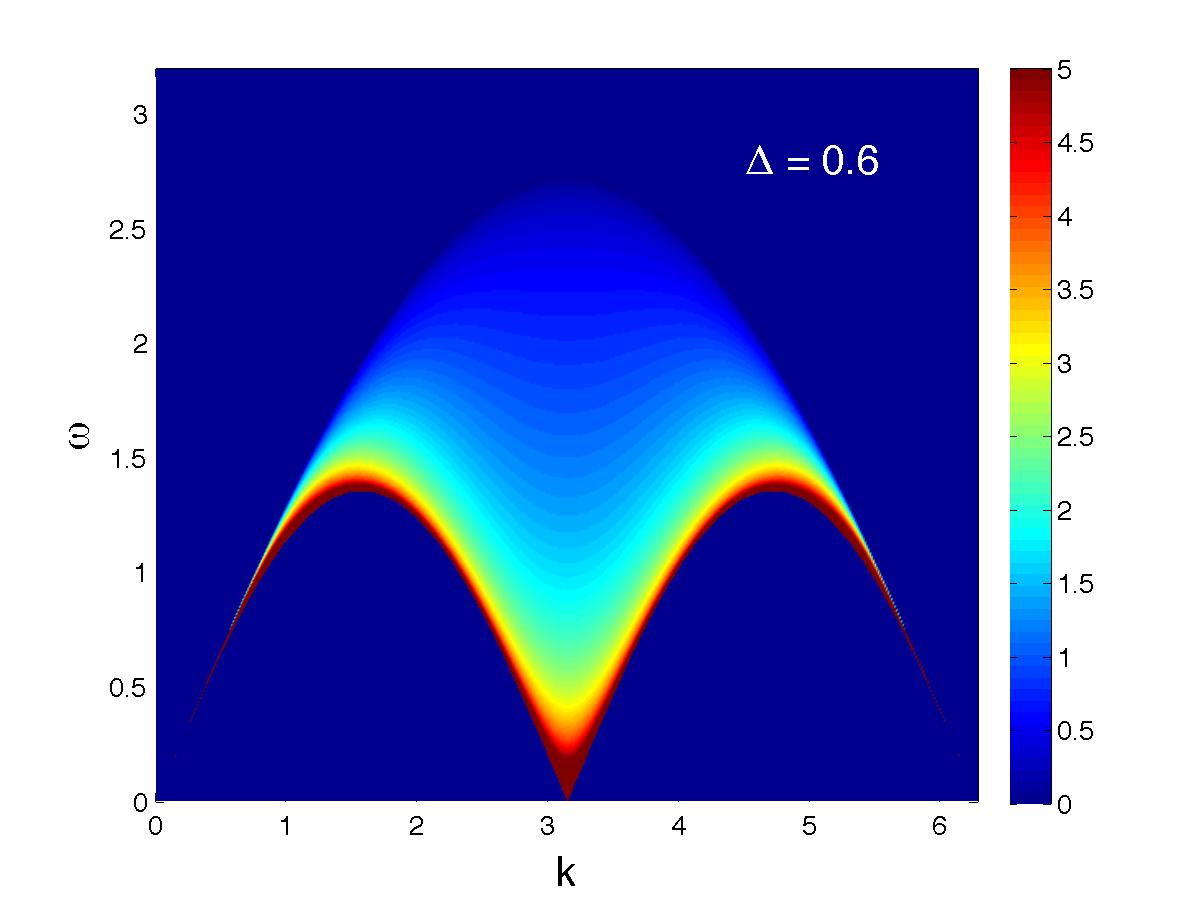}
&
\includegraphics[width=75mm]{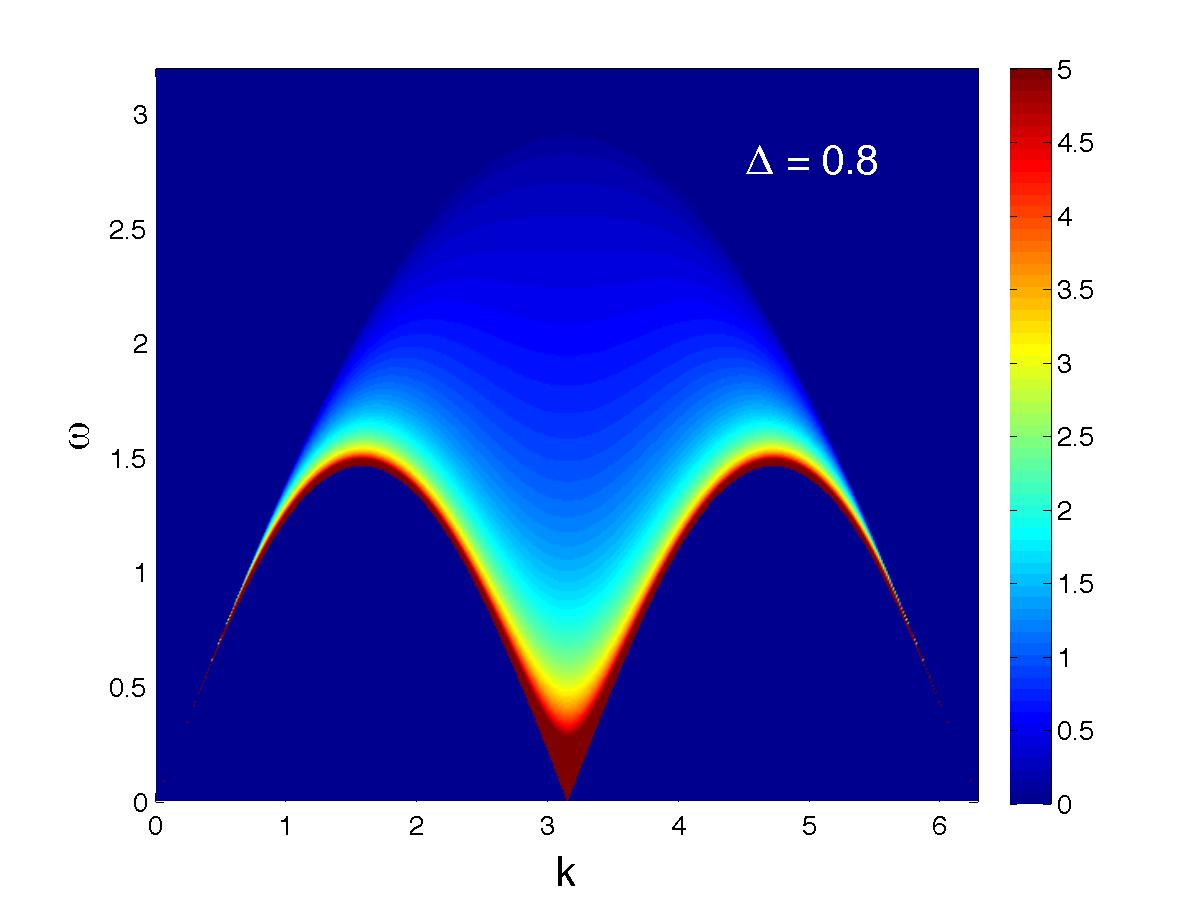}
\\
\includegraphics[width=75mm]{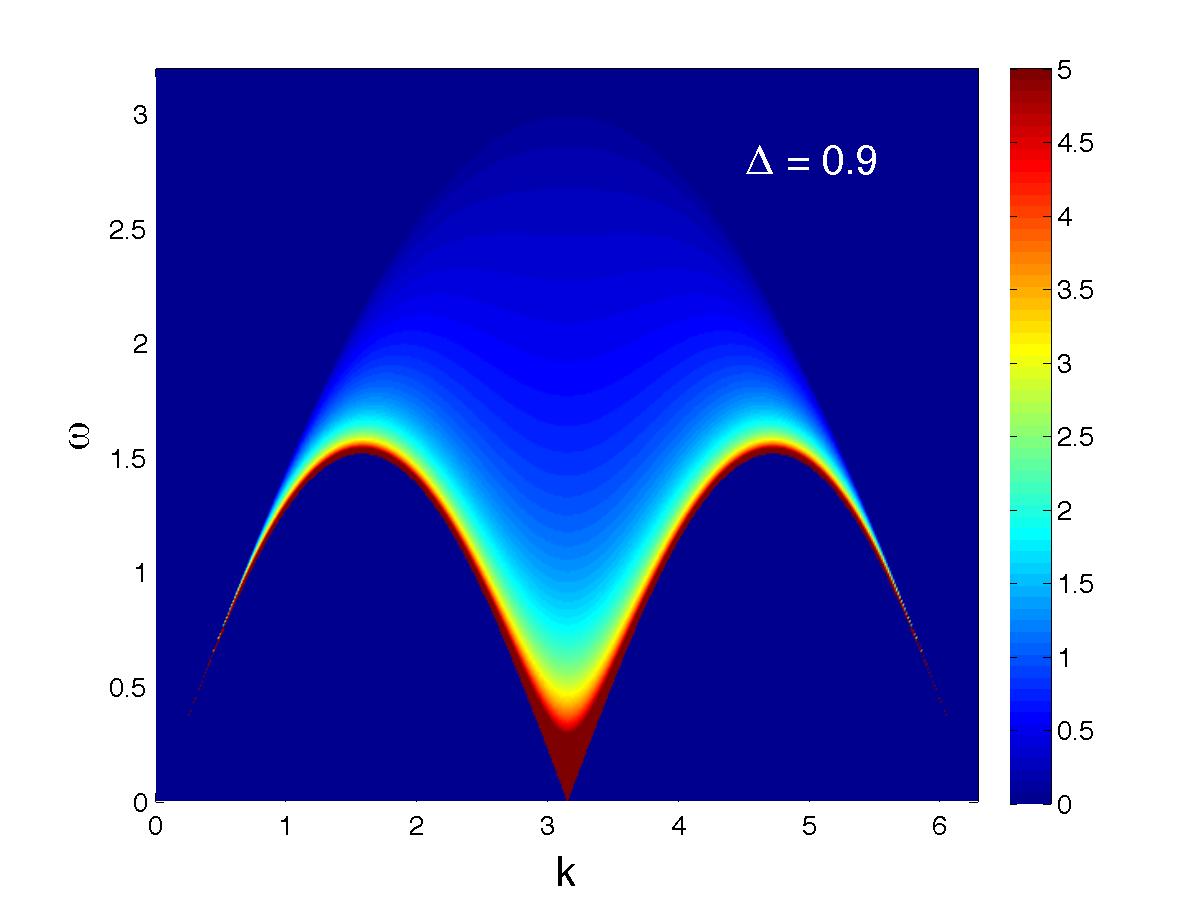}
&
\includegraphics[width=75mm]{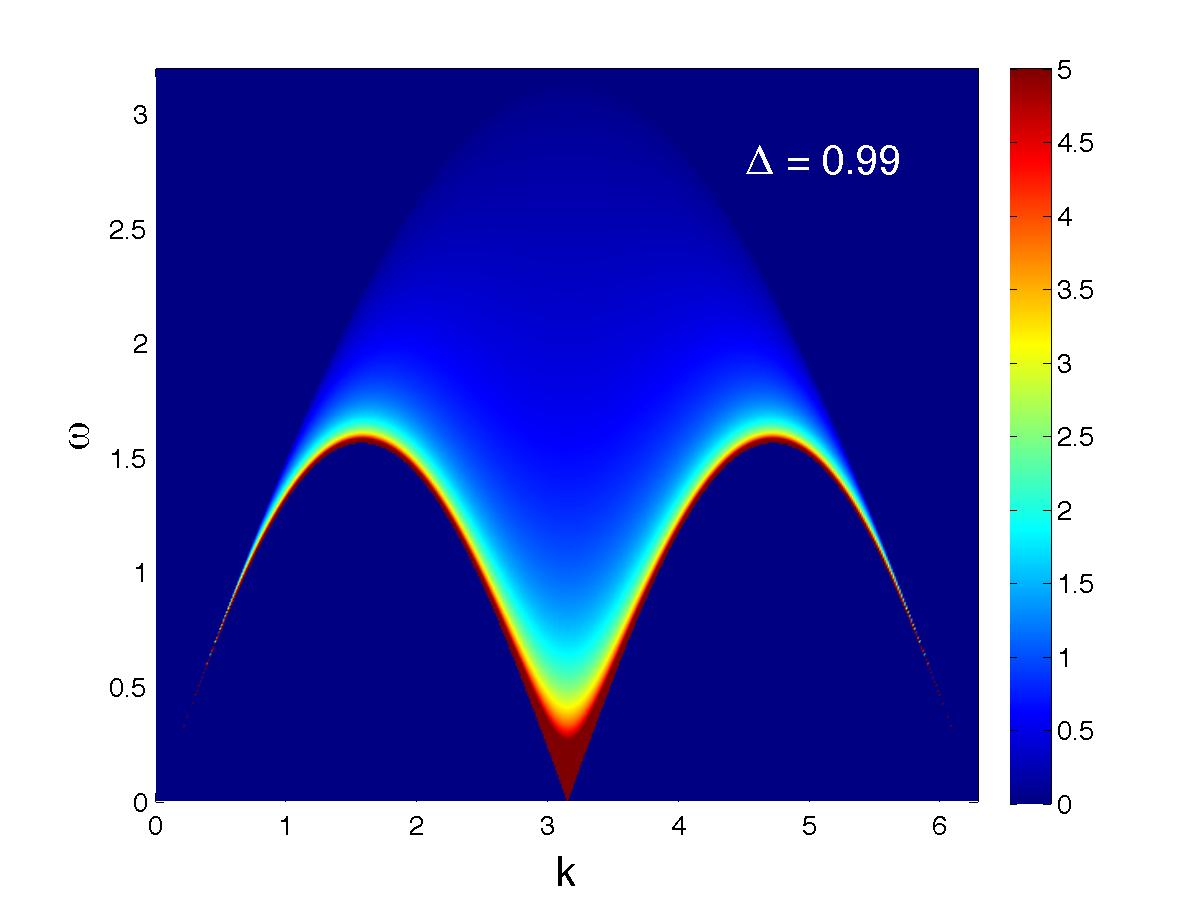}
\end{tabular}
\caption{Two-spinon part of the longitudinal structure factor of the infinite Heisenberg chain, 
for different values of the anisotropy parameter $\Delta$.  For $\Delta \rightarrow 0$, the
correlation follows the density of states, and has a square root singularity at the upper threshold for all values of momenta. Increasing the anisotropy shifts the weight progressively towards the lower boundary. The lower boundary becomes increasingly sharp as the $\Delta \rightarrow 1$ limit is approached.}
\label{fig:LSF}
\end{figure*}

{\it Results --}
In Fig.\ref{fig:LSF}, we give plots of the two-spinon part of the longitudinal
structure factor (\ref{S_2}) for values of $\Delta$ interpolating between
weak and strong coupling. A few striking things are worth mentioning 
concerning the influence of interactions on the two-spinon part of the correlations.
Most noticeably, upper threshold divergence disappears immediately 
upon turning interactions on. The correlation weight also starts flowing 
around the edges of the continuum, mostly via the wings at $k \simeq 0, 2\pi$ (see {\it e.g.} the $\Delta = 0.2$ plot),
and thereafter starts accumulating at the antiferromagnetic point $k = \pi$ (see the $\Delta = 0.4$ plot).
The lower threshold divergence starts carrying more weight from 
$\Delta \simeq 0.5$ onwards, and becomes increasingly sharp as one approaches the isotropic point.

Within the two-spinon continuum, so away from the thresholds, two things can be noticed.
First, the weight within the bulk of this continuum quickly changes shape as $\Delta$ is turned on:
from a pure $[\omega_{2,u} (k) - \omega]^{-1/2}$ form at $\Delta = 0$, it  
becomes almost uniform in frequency for $\Delta \simeq 0.2$; it then becomes a rapidly decreasing function
of frequency for higher interactions. Turning interactions on therefore leads to a remarkable 
collapse of correlation weight from {\it high} to {\it low} energies.

{\it Sum rules --}
To quantify the importance of the two-spinon contribution to the full structure factor, 
we use two useful sum rules, namely the integrated intensity
\begin{equation}
I^{zz} = \int_0^{2\pi} \frac{dk}{2\pi} \int_0^{\infty} \frac{d\omega}{2\pi} S(k,\omega) = 1/4,
\label{eq:sr}
\end{equation}
and the f-sumrule (at fixed momentum) \cite{1974_Hohenberg_PRB_10},
\begin{equation}
I^{zz}_1 (k) = \int_0^{2\pi} \frac{d\omega}{2\pi} \omega S (k,\omega) = - 2 X^x (1 - \cos k)
\label{eq:fsumrule}
\end{equation}
where $X^x \equiv \langle S^x_j S^x_{j+1} \rangle$ is the ground state expectation value of the in-plane exchange term.  
This can be obtained from the ground-state energy density $e_0$ \cite{1966_Yang_PR_150_2} and its derivative, 
namely $X^x = \frac{1}{2J} (1 - \Delta \frac{\partial}{\partial \Delta}) e_0$, with
\begin{equation}
e_0 = \frac{-J (\xi + 1)}{2\pi} \sin \left[\frac{\pi}{\xi \!+\! 1}\right] \int_{-\infty}^\infty \!\!\!dt \left( 1 - \frac{\tanh t}{\tanh [(\xi + 1) t]} \right).
\end{equation}
We provide the explicit values of the sum rule saturations coming from two-spinon contributions
in Table \ref{tbl:SR} (for the f-sumrule, the saturation is the same at all momenta).
The two-spinon states carry the totality of the correlation at $\Delta = 0$, and this remains
approximately true up to surprisingly large values of interactions $\Delta \sim 0.8$, above
which four, six, ... spinon states become noticeable.

\begin{table}[ht]
\begin{center}
\begin{tabular}{|c|c|c|c|c|c|c|}
\hline
$\Delta$ & $I^{zz}_{2sp}/I^{zz}$ & $I^{zz}_{1,2sp}/I^{zz}_1$ &  & $\Delta$ & $I^{zz}_{2sp}/I^{zz}$ & $I^{zz}_{1,2sp}/I^{zz}_1$ \\
\hline
0 & 1 & 1 & 
& 0.6 & 0.9778 & 0.9743 \\
0.1 & 0.9997 & 0.9997 &
& 0.7 & 0.9637 & 0.9578 \\
0.2 & 0.9986 & 9.9984 &
& 0.8 & 0.9406 & 0.9314 \\
0.3 & 0.9964 & 9.9959 &
& 0.9 & 0.8980 & 0.8844 \\
0.4 & 0.9927 & 0.9917 &
& 0.99 & 0.7918 & 0.7748 \\
0.5 & 0.9869 & 0.9849 &
& 0.999 & 0.7494 & 0.7331 \\
\hline
\end{tabular}
\end{center}
\caption{Sum rule saturations as a function of anisotropy:  two-spinon contribution to
the integrated intensity $I^{zz}$ (\ref{eq:sr}) and first frequency moment $I^{zz}_1$ (\ref{eq:fsumrule}).}
\label{tbl:SR}
\end{table}

{\it Threshold behaviour --}
The behaviour of the longitudinal structure factor in the vicinity of the excitation thresholds can be determined 
from the analytic expressions we have obtained, allowing us to make contact with and complement recent field
theory predictions \cite{2008_Pereira_PRL_100,2008_Cheianov_PRL_100}.

\paragraph{The structure factor near the upper threshold.} 
The upper threshold $\omega \rightarrow \omega_{2,u} (k)$ is approached by the limit $\rho \rightarrow 0$ 
as can be seen from (\ref{rhodef}).  A careful evaluation shows that the integral (\ref{I_rho})
then behaves according to
$I_\xi (\rho) ~\substack{\vspace{1mm} \\ \xrightarrow{\hspace{8mm}} \\ {\rho \rightarrow 0}} ~ -2\ln \rho + \mbox{O}(1).$
We thus have 
from (\ref{rhodef}) that the structure
factor vanishes as a square root,
\begin{equation}
S^{zz}_2 (k, \omega) 
~\substack{\vspace{1mm} \\ \xrightarrow{\hspace{14mm}} \\ {\omega \rightarrow \omega_{2,u}(k)}} ~
= f_u (\xi) (\sin\frac{k}{2})^{-7/2} \sqrt{\omega_{2,u}(k) - \omega}
\end{equation}
in which $f_u (\xi)$ is a momentum-independent function of anisotropy.
The anisotropy-independent square-root cusp at the threshold (for $0 < \Delta \leq 1$) 
confirms the field theory predictions \cite{2008_Pereira_PRL_100}, and at $\Delta \rightarrow 1$
matches the same limit known to apply for the $XXX$ case \cite{1997_Karbach_PRB_55}.
The prefactor we obtain here varies quickly with momentum, showing strong enhancement of the 
upper threshold singularity when taking the momentum towards the $k = 0, 2\pi$ zone boundaries (as can be seen in
Fig.\ref{fig:LSF}, most clearly at small anisotropies).  
For the $\Delta \rightarrow 0$ limit (so $\xi \rightarrow 1$), 
the $\cosh \frac{2\pi \rho}{\xi} + \cos \frac{\pi}{\xi}$ in the denominator of 
(\ref{S_2}) vanishes when $\rho \rightarrow 0$.
Overall, in this case one rather obtains a square-root divergence,
$S^{zz}_2 (k, \omega) 
~\substack{\vspace{1mm} \\ \xrightarrow{\hspace{14mm}} \\ {\omega \rightarrow \omega_{2,u}(k)}} ~
f_u (1) \frac{(\sin\frac{k}{2})^{-1/2}}{\sqrt{\omega_{2,u}(k) - \omega}},$ 
which follows the singularity of the density of states (the matrix elements are then energy independent).
This discontinuous in $\Delta$ threshold exponent behaviour is also consistent with field theory \cite{2008_Pereira_PRL_100}.
We notice further that the momentum dependence of the prefactor is changed to a much weaker one than that at $\Delta \neq 0$.

\paragraph{The structure factor near the lower threshold.} 
The limit 
$\omega \rightarrow \omega_{2,l} (k)$ is obtained via $\rho \rightarrow \infty$.
Evaluating (\ref{I_rho}) yields 
$I_\xi (\rho) 
~\substack{\vspace{1mm} \\ \xrightarrow{\hspace{10mm}} \\ {\rho \rightarrow \infty}} ~
-\pi \left(1 + \frac{1}{\xi}\right) \rho + \mbox{O}(1).$
The structure factor then obeys
\begin{equation}
S^{zz}_2 (k, \omega) 
~\substack{\vspace{1mm} \\ \xrightarrow{\hspace{14mm}} \\ {\omega \rightarrow \omega_{2,l}(k)}} 
= f_l (\xi) \frac{|\sin k|^{-\frac{1}{2} (1 - \frac{1}{\xi})} (\sin\frac{k}{2})^{-\frac{2}{\xi}}}
{[\omega - \omega_{2,l} (k)]^{\frac{1}{2}(1 - \frac{1}{\xi})}},
\end{equation}
where $f_l (\xi)$ is again a momentum-independent function of anisotropy.
The singularity exponent reproduces an early conjecture \cite{1981_Mueller_PRB_24,1981_Mueller_JPC_14} and
field theory predictions \cite{2008_Pereira_PRL_100,2008_Cheianov_PRL_100};
the momentum-dependent part of the prefactor shows an even more complicated behaviour than that of the
upper threshold, being enhanced (though differently) both at the zone boundaries $k = 0,2\pi$ as well as near 
$k = \pi$.
As a final detail, the $\Delta \rightarrow 0$ limit (so $\xi \rightarrow 1$) yields the expected behaviour,
$S^{zz}_2 (k, \omega) 
~\substack{\vspace{1mm} \\ \xrightarrow{\hspace{14mm}} \\ {\omega \rightarrow \omega_{2,l}(k)}} ~
O(1)$. 

{\it Conclusions --}
In this paper, we have tracked how the spinon excitations in Heisenberg antiferromagnets contribute to 
the longitudinal spin structure factor (\ref{eq:SZZ}), as a function
of anisotropy ({\it i.e.} interaction).
We have obtained the two-spinon part of this correlator exactly in the zero-field, infinite-size chain
throughout the gapless antiferromagnetic regime,  
by exploiting the vertex operator approach to
express states and correlators in a purely algebraic language.
Our results provide an exact lower bound for and an extremely accurate description of the full 
correlator (as shown by sum rule saturations) 
throughout the observable excitation continuum ({\it i.e.} not only at low energies or near thresholds), 
provide a resilient check for alternate methods 
and give a nonperturbative derivation of the threshold exponents
obtained from field theory while complementing these with exact prefactors. The precise functional form
we have obtained also allows us to determine the region of validity of the threshold behaviour; we will
address this and further issues in future work \cite{CauxKonnoSorrellWeston_TBP}.

{\it Acknowledgements --}
J.-S. C. acknowledges support from the FOM foundation of the Netherlands. 
H.K. was supported in part by Grant-in-Aid for Scientific Research (C) 22540022.
M.S. acknowledges the Australian Research Council (ARC) for financial support.
The authors are grateful to
L. Frappat and E. Ragoucy as organizers of the RAQIS conferences, during which this work was initiated.

\bibliography{/Users/jscaux/WORK/BIBTEX_LIBRARY/JSCAUX_papers,/Users/jscaux/WORK/BIBTEX_LIBRARY/OTHERS_TBP,/Users/jscaux/WORK/BIBTEX_LIBRARY/BOOKS,/Users/jscaux/WORK/BIBTEX_LIBRARY/1926-1930,/Users/jscaux/WORK/BIBTEX_LIBRARY/1931-1935,/Users/jscaux/WORK/BIBTEX_LIBRARY/1936-1940,/Users/jscaux/WORK/BIBTEX_LIBRARY/1941-1945,/Users/jscaux/WORK/BIBTEX_LIBRARY/1946-1950,/Users/jscaux/WORK/BIBTEX_LIBRARY/1951-1955,/Users/jscaux/WORK/BIBTEX_LIBRARY/1956-1960,/Users/jscaux/WORK/BIBTEX_LIBRARY/1961-1965,/Users/jscaux/WORK/BIBTEX_LIBRARY/1966-1970,/Users/jscaux/WORK/BIBTEX_LIBRARY/1971-1975,/Users/jscaux/WORK/BIBTEX_LIBRARY/1976-1980,/Users/jscaux/WORK/BIBTEX_LIBRARY/1981-1985,/Users/jscaux/WORK/BIBTEX_LIBRARY/1986-1990,/Users/jscaux/WORK/BIBTEX_LIBRARY/1991-1995,/Users/jscaux/WORK/BIBTEX_LIBRARY/1996-2000,/Users/jscaux/WORK/BIBTEX_LIBRARY/2001-2005,/Users/jscaux/WORK/BIBTEX_LIBRARY/2006-2010}

\end{document}